\documentclass[a4paper]{article}
\usepackage{amssymb}                                          %%
\usepackage{amsmath}                                          %%
\usepackage{amssymb}

\usepackage{graphics}
\usepackage{graphicx}
\usepackage{wrapfig,epsfig,epsf}
\usepackage{bm}    %%%% boldface in math mode

\makeatletter
\renewcommand\@biblabel[1]{#1.} %%% refs 1.
\makeatother

\usepackage[left=2.5cm,right=2.5cm,top=4cm,bottom=2.7cm]{geometry}

\def\beq#1{\begin{equation}\label{#1}}
\def\eeq{\end{equation}}
\def\beqa#1{\begin{eqnarray}\label{#1}}
\def\eeqa{\end{eqnarray}}

\def\eqn#1{~(\ref{#1})}
\def\myfrac#1#2{\left(\frac{#1}{#2}\right)}

\def\comment#1{\relax}

\begin{document}

\title{Spins of black holes in coalescing compact binaries}

\author{K.A. Postnov$^{1,2,3}$, A.G. Kuranov$^{1,4}$, N.A. Mitichkin$^{1,2}$\\
$^1$Sternberg Astronomical Institute\\
M.V. Lomonosov Moscow State University,
Universitetskij pr., 13, 119234, Moscow, Russia\\
$^2$Faculty of Physics MSU,\\
Leninskie Gory 1, 199991, Moscow, Russia\\
%Faculty of Physics, M.V. Lomonosov Moscow State University,\\ Leninskie Gory, 1, 119991, Moscow, Russia\\
$^3$Faculty of Physics, Novosibirsk State University\\
Pirogova 2, 630090, Novosibirsk, Russia\\
%Department of Physics, Novosibirsk State University, \\Pirogova 2, 630090, Novosibirsk, Russia\\
$^4$ Russian Foreign Trade Academy,  4a Pudovkin str., 119285 Moscow, Russia
}
\maketitle

\section*{Abstract}

Modern astrophysical methods of determination of spins of rotating stellar-mass black hole in close binaries and of supermassive black hokes in active galactic nuclei are briefly discussed. Effective spins of coalescing binary black holes derived from LIGO/Virgo gravitational wave observations are specially addressed.  We consider three types of coalescing binaries: double black holes, black hole -- neutron star binaries and primordial double black holes. 
The effective spins of coalescing astrophysical binary black holes and black holes with neutron stars are calculated for two plausible models of black hole formations from stellar core collapses (without or with additional fallback from stellar envelope) taking into account the stellar metallicity and star formation rate evolution in the Universe. 
The calculated distributions do not contradict the reported LIGO/Virgo observations. The effective spins of primordial coalescing stellar-mass black holes can reach a few per cent due to accretion spin-up in the cold external medium.

\section{Introduction}

\subsection{The Kerr metric}

In A. Einstein's general relativity (GR), the structure of space-time around a stationary rotating black hole (BH) in a vacuum is fully described by the axially symmetric steady Kerr metric   
 \cite{1963PhRvL..11..237K}. 
 The parameters of the metric include the BH mass 
 $M$ and angular momentum $J$ which can be conveniently expressed  in units  $a=J/(GM/c)$. In the Boyer-Lindquist coordinates  \cite{1967JMP.....8..265B} minimizing the number of non-diagonal metric elements, the Kerr metric can be written in the form \footnote{Here we use geometrical units $G=c=1$ unless stated otherwise.} :
\begin{eqnarray}
    \label{e:Kerr}
ds^2=&-\bigg(1-\frac{2Mr}{\Sigma}\bigg)dt^2-\bigg(\frac{4Mar\sin^2\theta}{\Sigma} \bigg) dtd\phi
+\frac{\Sigma}{\Delta}dr^2+\Sigma d\theta^2 \nonumber\\
&+\bigg(r^2+a^2+\frac{2Ma^2r\sin^2\theta}{\Sigma}\bigg)\sin^2\theta d\phi^2\*\
&\Delta\equiv r^2-2Mr+a^2,\quad \Sigma\equiv r^2+a^2\cos^2\theta\nonumber
\end{eqnarray}
    
The richness of mathematical structure and physical effects in the Kerr metric is described in dedicated texts (see, e.g., 
 \cite{2009kesp.book.....W}). The condition for the horizon 
($g_{rr}\to \infty$) and absence of a 'naked' singularity in a rotating BH require  $-M\le a\le M$. Measuring the mass in units of the BH mass $M$ renders the Kerr parameter
$a$ dimensionless,  $|a^*|\le 1$. Below we will use the dimensionless parameter  $a^*$ and call it the 'BH spin'.

For $|a|\to 0$ the Kerr metric (\ref{e:Kerr}) transforms into the Schwarzschild metric, and at large distances the Kerr metric \textit{asymptotically} \footnote{Unlike a non-rotating spherically symmetric body around which the space time is exactly described by the Schwarzschild metric (Birkhoff theorem).} turns into a metric for steadily rotating body in a vacuum that was derived from Einstein equations in the weak-field limit in 1918 by Lense and Thirring    \cite{1918PhyZ...19..156L}. From the astrophysical point of view, any celestial body rotates and therefore the collapse of massive stars should generally result in the formation of rotating BHs with non-zero parameter  $a$.

An important parameter we will use below is the notion of last marginally stable orbit of test particles (or innermost stable circular orbits, ISCO) in the Kerr metric. These orbits determine the inner boundary of an accretion disc around a rotating BH and are probed by astrophysical observations.   
In the equatorial orbit $\theta=\pi/2$, the ISCO radius  $r_\mathrm{ISCO}$ is determined by the BH spin $a$ \cite{1972ApJ...178..347B}: for a Schwarzschild BH ($a=0$)   $r_\mathrm{ISCO}=6M=3r_g$  (three gravitational radii $r_g=2M\simeq 3 \hbox{km} (M/M_\odot)$) \cite{1949ZhETF..19..951K}. For $a>0$, i.e. in the case of corotating particles, the ISCO radius decreases as $r_\mathrm{ISCO}\approx 3r_g-(4\sqrt{6}/3) a$ at $a\ll M$ and in the case $a=M(1-\delta)$, $\delta \ll 1$ approaches $r_\mathrm{ISCO}\simeq (r_g/2)(1+(4\delta)^{1/3})$, being above the outer BH horizon  ЧД $r_+\approx (r_g/2)(1+(2\delta)^{1/2})$ \cite{1970Natur.226...64B}. For example, for an accretion disc in a close binary system or in a galactic nucleus $a\approx 0.9981$ (the Thorne limit determined by the balance of the angular momentum balance for an accreting BH in the photon field of the disc) \cite{1974ApJ...191..507T}, the ISCO radius is  $r_\mathrm{ISCO}(0.998)\approx 1.23M$.

Thus, both mass $M$ and spin $a^*$ of a BH are the most important parameters that can be measured or estimated from astrophysical observations. These observations are different for stellar-mass BHs ($\sim 3-60 M_\odot$) in close binaries and supermassive BHs (SMBHs) with $M>10^6 M_\odot$ in galactic nuclei.

%\subsection{Методы оценки спина астрофизических черных дыр}

Methods of BH parameter estimations from astronomical observations have been largely discussed in the literature (see, for example, reviews 
\cite{Cherepashchuk:2014,Cherepashchuk:2016} and referenced therein).
The most reliable mass estimates of stellar-mass BHs are obtained from dynamical measurements of the motion of the secondary companion (optical star) in close binary systems (CBS). Orbits in such systems are almost circular. In this case, the radial velocity amplitude measured from spectroscopic observations
 $K_v=V_v\sin i$ ($i$ is the binary inclination angle to the line of sight) and the binary orbital period  $P_b$ enable the construction, using formulas of the classical two-body problem, of a combination with the dimension of mass, the so-called 'mass function' of the invisible component with mass  $M_x$, which for the circular orbit reads 
\beq{e:f(M)}
f(M_x)=\frac{K_v^3P_b}{2\pi G}=\frac{M_x\sin^3 i}{(1+q)^2}\,,
\eeq
where $q=M_v/M_x$ is the mass ratio of the visible to invisible component of the binary. Clearly, the mass function sets the lower mass limit of the unseen component because the determination of the orbital binary inclination requires independent measurements
\footnote{The binary orbit inclination cannot be obtained from spectroscopic measurements using Newtonian dynamics. However, this becomes possible in very close binaries with neutron stars -- pulsars using relativistic effects \cite{1975SvAL....1....2B}.}. It is accepted to consider the condition 
$M_x\gtrsim 3 M_\odot$ for the mass of the unseen component in a CBS as the necessary signature of a BH
\cite{1974PhRvL..32..324R,2016PhR...621..127L}. This mass limit of the unseen relativistic component without evidence of a solid surface is met in two dozens of close X-ray binaries  \cite{2006ARA&A..44...49R,Cherepashchuk:2016}. 

The advent of gravitational-wave (GW) astronomy
\cite{LIGO-PRL} offered a  novel possibility to estimate the masses and spins of BH components in coalescing binary systems observed by LIGO/Virgo GW interferometers \cite{LIGOO2,Cherepashchuk:2016b,Raittse:2017}.

\subsection{Black hole spins in X-ray binaries and active galactic nuclei}

Measurements of BH spins is a much more difficult task than measurements of BH masses and are model-dependent. In X-ray binaries, the matter from gravitationally captured stellar wind of the optical star of early spectral class (e.g., in Cyg X-1) or from Roche lobe overflow by the optical star   
(e.g., in GRS 1915+105, A0620-00, etc.) forms an accretion disc around the BH. The viscous friction in the differentially rotating disc heats up the gas to very high temperatures, and the matter loses its angular momentum and gradually diffuses in the thermal time scale towards the BH. The inner radius of the accretion disc is determined by the innermost stable circular orbit around the BH $r_\mathrm{ISCO}$ inside which particles freely fall into the BH. The structure and properties of stationary accretion discs around BHs are determined by the mass accretion rate  $\dot M$ through the disc, physical parameters of the gas (radiation opacity, viscosity coefficient, etc.) and by the boundary conditions at the inner radius \cite{1972AZh....49..921S,1973A&A....24..337S,1973blho.conf..343N,2018ASSL..454..115Z}. 

Thus, measurements of spectral and time characteristics of the emission from accretion discs around BHs enables indirect estimation of their masses and spins. The modern review of such observations and the list of sources can be found in \cite{2018arXiv181007041N}.

The spin of BHs in X-ray binaries and galactic nuclei can be estimated in different ways (see \cite{2019NatAs...3...41R} for the recent review and critical comments in \cite{2019NatAs.tmp..241L}). 

\textbf{The continuum spectrum of accretion disc}.
Through the boundary condition at the inner disc radius the BH spin affects the form of the thermal spectral continuum radiated by the accretion disc. 
Numerous model parameters (the type of the accretion disc, BH mass, distance to the source, inclination to the line of sight, parameters of the hot corona above the disc, etc.), however, do not allow precise measurements of BH spins (see Table 2 for X-ray binaries in  \cite{2018arXiv181007041N}). 

\textbf{Fluorescent iron line spectroscopy}.
Spins of accreting BHs can also be estimated from the spectroscopy of the
K-$\alpha$ iron line (at the energy from 6.4 keV for the neutral iron to 
6.9 keV for hydrogen-like iron  Fe XXVI). 
This line emerges during the reflection of hard X-ray emission that can be generated in a hot scattering corona above the disc from relatively cold gas in the inner parts of the disc ($<10^7$ K) \cite{1991MNRAS.249..352G}. 
The reflected line has an equivalent width of $\sim 150$~eV. Due to relativistic effects at the inner disc radius, the line profile is red-shifted and gets a width of$\sim 1$~keV \cite{1991ApJ...376...90L}. 
Depending on the line of sight and the assumed emission model, the blue wing of the line sharply drops at the gravitationally red-shifted frequency. The fluorescent iron line from the inner parts of the accretion disc around a SMBH was first observed in the galaxy   MCG-6-30-15 \cite{1995Natur.375..659T}.
For SMBHs in galactic nuclei, the spin determination from   
К-$\alpha$ iron line spectroscopy is the only available (besides the direct EHT observations of the BH shadow in the nucleus of M87 galaxy, see below).
The profiles of the fluorescent 
 K-$\alpha$ and L-$\alpha$ iron lines have been accurately measured in spectra of several sources, in particular in Seyfert galaxy  1H0707-495 \cite{2009Natur.459..540F}.
The list of SMBHs with spins calculated by the iron line spectroscopy can be found, for example, in Table 3 of review  \cite{2018arXiv181007041N}.
Unlike continuum spectral fitting, the analysis of the fluorescent iron line profile formed by the reflection from the disc does not require the knowledge of the BH mass, the distance to the source and the disc inclination angle to the observer. Note that both methods yield consistent estimates of the parameter $a^*$ for accreting BHs in X-ray binaries (see Table 2 in  \cite{2018arXiv181007041N}). 

\textbf{X-ray QPOs}.
Observations of quasi-periodic oscillations (QPO) of the X-ray flux
offer yet another method to estimate spins of accreting BHs  
\cite{2004astro.ph.10551V}. 
In BH sources, QPOs have been observed at different frequencies, including low-frequency QPOs (fraction of Hz -- dozen Hz) and high-frequencies QPOs (several 100 Hz). High-frequency QPOs are close to some characteristic oscillations for stellar-mass BHs: the Keplerian $f_K$, radial $f_r\sim f_K$ and vertical epicyclic $f_\theta\sim f_K$ frequencies \cite{1999ApJ...524L..63S,2001A&A...374L..19A}: 
\beq{e:fQPO}
f_K(r)\simeq 220\,\hbox{Hz} \myfrac{10M_\odot}{M}\myfrac{6M}{r}^{3/2}\bigg[1\pm a^*\myfrac{M}{r}^{3/2}\bigg]^{-1}\,.
\eeq

Applying this model to X-ray QPOs observed by the RXTE telescope enabled high-precision estimations of masses and spins of BHs in 
GRO J1655-40 ($M = (5.31 \pm 0.07)M_\odot$, $a^* = 0.290 \pm 0.003$) \cite{2014MNRAS.437.2554M} and XTE J1550-564  ($a^* = 0.34\pm 0.01$) \cite{2014MNRAS.439L..65M}), which coincided with spectroscopic estimates. Note also that the observed spectral correlations of X-ray QPOs in accreting BHs can be used to independent measurements of BH masses  \cite{2009ApJ...699..453S}.

\textbf{Observations of SMBH in M87 by the  Event Horizon Telescope (EHT)}.
Recently, first results of 1.3-mm VLBI observations of SMBH in active nucleus of M87 galaxy ($M=(6.5\pm 0.7) \times 10^9M_\odot $) with record high angular 10-microarcsec resolution carried put by the EHT dishes were reported \cite{2019ApJ...875L...1E}. 
Relativistic MHD modelling of the observed asymmetric radio brightness in the visible 'photon ring' from optically thin hot plasma around this SMBH with an account of relativistic effects of photon propagation in gravitational field of a Kerr BH enabled the BH spin estimate $a^*\approx 0.5$ or $a^*\simeq 0.94$ \cite{2019ApJ...875L...5E} 
(see also an independent analysis \cite{2019arXiv190607171D} yielding $a^*\simeq 0.75\pm 0.15$ ). 
Note that models of a Schwarzschild BH are rejected by these observations as well as by the presence of relativistic jet from the M87  nucleus with a power of $P_j\gtrsim 10^{42}$~erg/s. An independent estimate of the BH spin in M87 $a^*=0.9\pm 0.1$ was also obtained in paper \cite{2019arXiv190407923T} from the analysis of the 'twisted light' in the Kerr metric. 

Thus, observations of accreting BHs in X-ray binaries and galactic nuclei suggest a fairly high spin in these BHs. This is apparently related to a prolonged accretion of matter and the history of SMBH mass growth in galactic nuclei in due course of galactic mergings. 

\section{Spins of coalescing binary black holes}
\label{s:bhbh}
\subsection{Effective spins of  LIGO/Virgo coalescing binaries}

LIGO/Virgo observations of coalescing binary BHs 
\cite{LIGOO2} provide new independent information on masses and spins of coalescing binary BHs. In the quadrupole approximation, the form and amplitude of GW signal from inspiraling compact objects depend of the combination of masses 
 $M_1$ and $M_2$ called chirp mass: ${\cal M}\equiv (M_1M_2)^{3/5}/M^{1/5}=M_1(q^2(1+q))^{-1/5}$, where $q=M_1/M_2$ is 
 the binary mass ratio and $M=M_1+M_2$ is the total mass of the system. The chirp mass is the most precise parameter that can be determined from GW observations (see \cite{LIGOO2}). 
 The spins of the binary components are much more difficult to estimate. However, the analysis of the form of the inspiraling GW signal enables the measurement of the so-called 'effective spin' of the coalescing binary, a mass-weighted combination of components' spin projections on the orbital angular momentum:
\beq{e:chieff}
\chi_{\mathrm{eff}}=\frac{M_1a_1^*\cos\theta_1+M_2a_2^*\cos\theta_2}{M}\,,
\eeq{}
where $\theta_i$ is the angle between the i-th component's spin and the binary orbital angular momentum. 

Most of the detected LIGO/Virgo binary BHs (but two sources, 
GW151226 and GW170729) have been found to have near-zero effective spin (within the measurement errors)  \cite{LIGOO2}. 
At first glance, this fact seems unusual because collapsing massive stars should have rotating cores \cite{2017hsn..book..601M}
leading to the formation of BH with non-zero spins. Rapidly rotating BHs from stellar core collapses have been thought to be 'central engines' generating narrow-beamed relativistic jets observed as long gamma-ray bursts \cite{2006ARA&A..44..507W}. Therefore, it is interesting to understand whether is it possible to get close to zero effective spins of coalescing binary BHs formed in the standard astrophysical evolutionary channel from massive binary field stars
\cite{2001PhyU...44R...1G,2014LRR....17....3P,2016Natur.534..512B}.

\subsection{Model assumptions}

It is difficult to measure the rotation of stellar cores from observations. The theory of evolution of rotating stars involves many parameters to describe to rotation of the stellar core, including, for example, the hypothesis about core-envelope coupling by the internal magnetic field
\cite{2002A&A...381..923S}, internal inertial gravitational waves
\cite{2015ApJ...810..101F} and other physical mechanisms of the angular momentum transport (see 
\cite{2017hsn..book..601M} for more detail).
To describe the complicated evolution of the core of a massive star, in paper  \cite{2016MNRAS.463.1642P} a semi-phenomenological approach was suggested, in which the star was split into two parts: the core and the envelope, whose coupling was described by one effective parameter -- the characteristic time of the angular momentum transfer 
$\tau_c$. In this two-zone model, the change of the angular momentum of the core is written as 
\beq{e:c-e}
\frac{d\bf{J_c}}{dt}=-\frac{I_cI_e}{I_c+I_e}\frac{\bf{\Omega_c}-\bf{\Omega_e}}{\tau_c}\,,
\eeq
where $I_c$, $\Omega_c$ and $I_e$, $\Omega_e$ are the moments of inertia and angular rotational velocities of the core and envelope, respectively. The paper  \cite{2016MNRAS.463.1642P} 
showed that to describe the distribution of rotation periods of young pulsars as derived from observations, the characteristic coupling time should be of the order of 
$\tau_c\simeq 5\times 10^5-10^6$~years, i.e. should approximately 
coincide with the evolutionary time of a massive star after the main sequence.

If a BH is born in a massive binary system, the rotation of the stellar core prior to the collapse will be also affected by the tidal coupling between the stellar envelope and the orbital motion of the secondary component. Taking into account the results  obtained in 
\cite{2016MNRAS.463.1642P}, in paper \cite{2019MNRAS.483.3288P} 
the results of calculations of the effective spins of coalescing binary BHs were presented for different formation channels with an account of both metallicity and star formation history (dependence of redshift) in galaxies. These calculations adopted the standard theory of evolution of massive close binaries 
\cite{2014LRR....17....3P} added with the treatment of rotational evolution of stellar cores using the coupling \eqn{e:c-e}. The key unknown element of the double BH formation from massive binaries is the common envelope (CE) stage. This stage has been treated using the efficiency parameter $\alpha_\mathrm{CE}$, which is the fraction of orbital energy transferred to the stellar envelope during binary imspiral in the common envelope:
$\Delta E_{env}=\alpha_\mathrm{CE}\Delta E_{orb}$  ($E_{env}$ is the binding energy of the envelope with stellar core) \cite{1984ApJ...277..355W,1984ApJS...54..335I}. 

Poorly known physics of the BH formation in the end of massive star evolution was parametrized by two models. In the first model, the entire stellar C-O core formed after the main sequence was collapsed into BH 
$M_{BH}=0.9 M_{CO}$ (with an account of a 10\% gravitational mass defect) so that the total mass of a binary BH was 
$M=0.9(M_{CO,1}+M_{CO,2})$. 
The BH angular momentum in this model was set equal to that of the C-O core: $J_{BH}=J_{CO}$.

In the second model, part of stellar envelope above the C-O core
$\Delta M_{fb}$ was assumed to fallback onto a BH collapsed from the iron core $M_{Fe}$ (the fallback model). Here the BH mass was calculated as in \cite{2012ApJ...749...91F}. Thus, in this model the BH mass was calculated as $M_{BH}=0.9(M_{Fe}+\Delta M_{fb}$) with $\Delta M_{fb}=\max(M_{BH}-0.9 M_{CO},0)$. 
The BH angular momentum changed due to the fallback from the rotating envelope
$J_{BH}=J_{CO}+\Delta J_{fb}$, where $\Delta J_{fb}=j_{fb}\Delta M_{fb}$. 
The specific angular momentum of the rotating envelope matter accreting onto the BH was assumed to be equal to 
$j_{fb}=\delta M_{BH}$ with the coefficient $\delta=2$
(the mean value between the specific angular momentum of particles on ISCO for a Schwarzschild BH, $\delta=2\sqrt{3}$, and an extreme Kerr BH,  $\delta=2/\sqrt{3}$).

To calculate the effective spin of coalescing BHs the angle between the BH spin vector and the orbital angular momentum should be specified (see equation \eqn{e:chieff}). 
This misalignment was calculated for two limiting cases: a) for initially aligned spins of main-sequence stars with  the orbit and b) for random initial orientation of stellar spins. In the first case, non-zero angles 
 $\theta_i$ prior to the coalescence can be expected only if some additional kick is imparted to the BH during the collapse (as in the case of neutron stars, see the discussion in   \cite{2014LRR....17....3P}).  
 For randomly misaligned initial spins of the binary components the angle $\theta_i$ can be arbitrary (see \cite{2019MNRAS.483.3288P} for more detail).

\subsection{Results of calculations}
\begin{figure}
\includegraphics[width=0.8\textwidth]{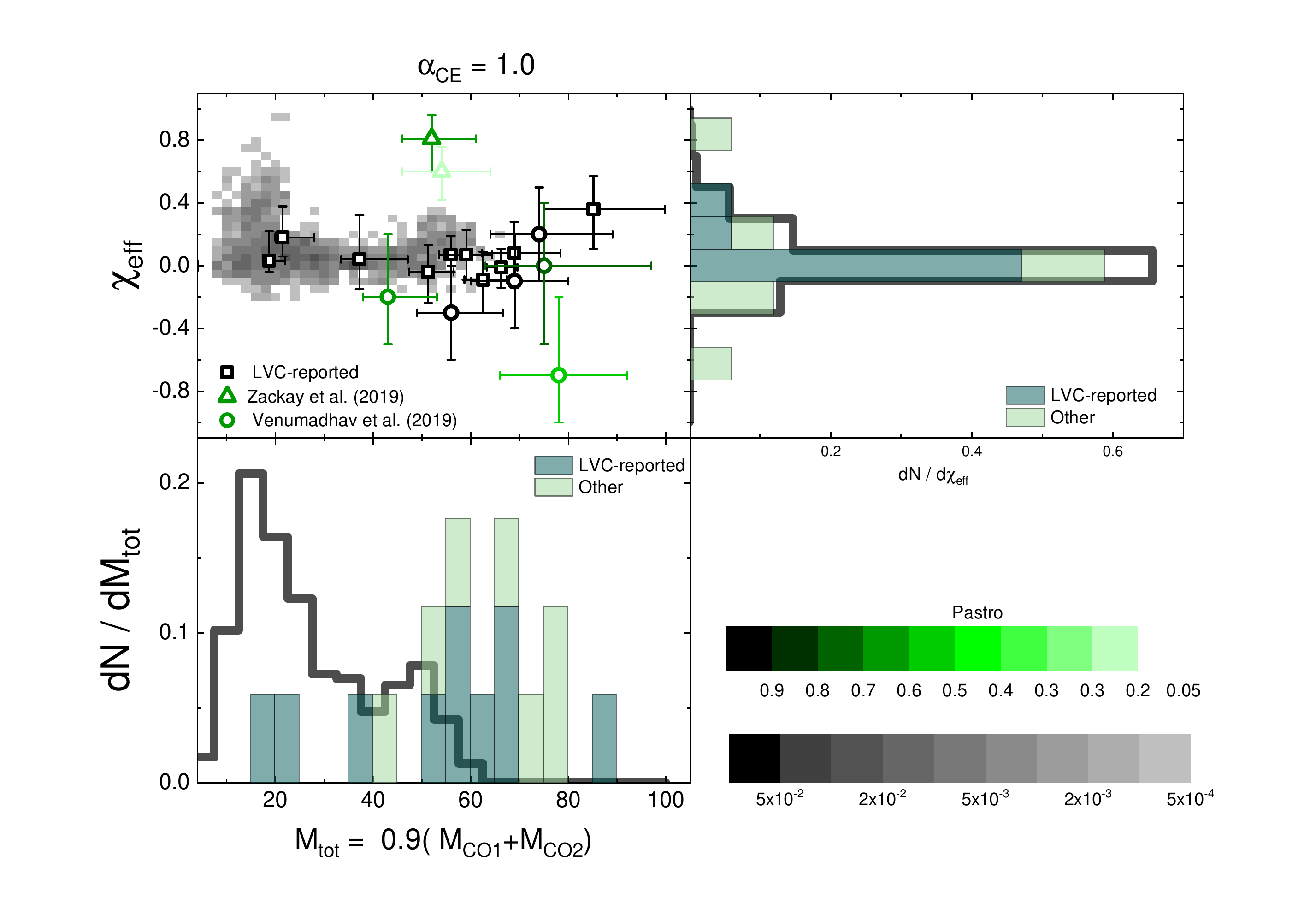}
\vfill
\includegraphics[width=0.8\textwidth]{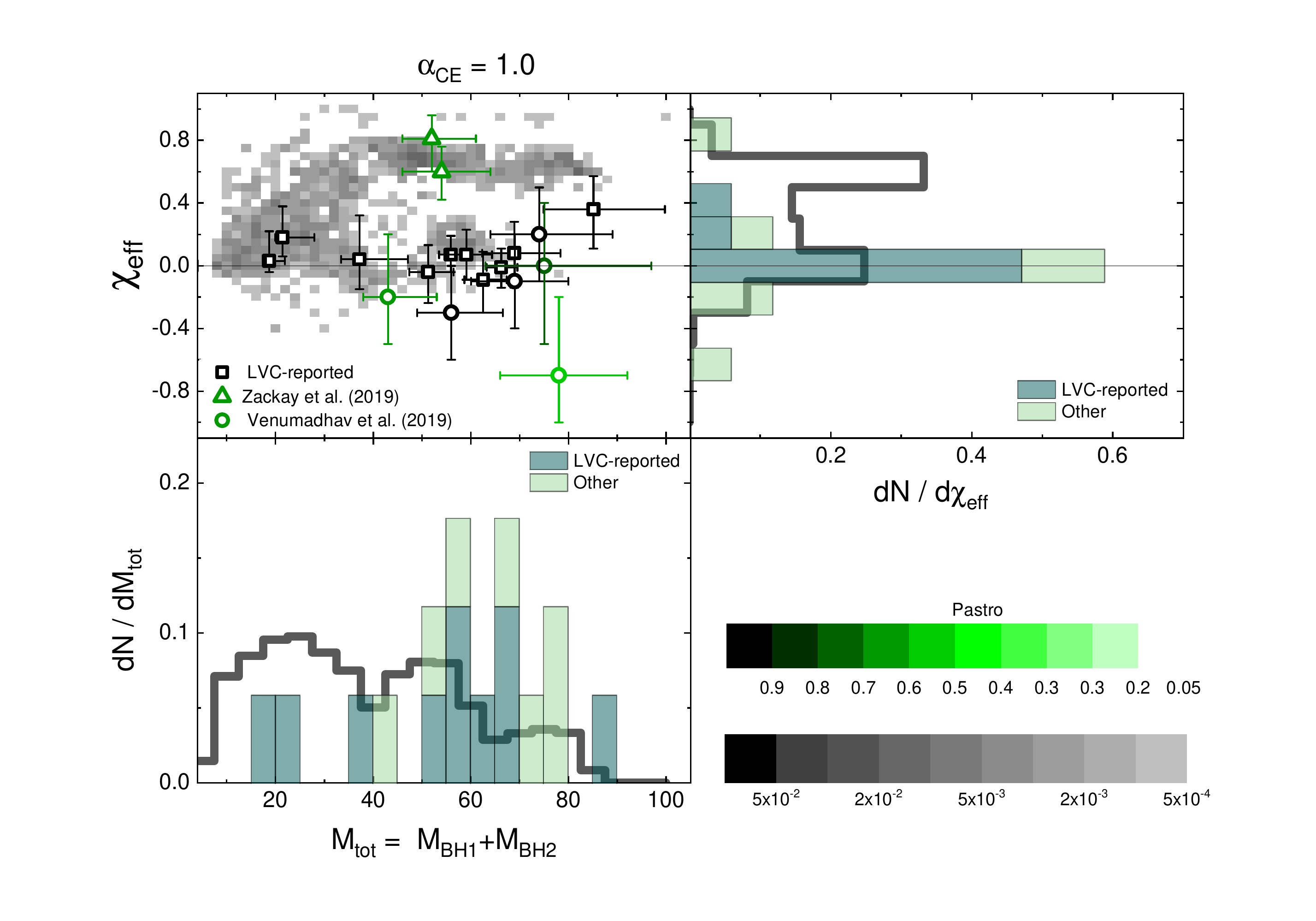}
\caption{Model total mass  $M_{tot}$ -- effective spin $\chi_\mathrm{eff}$ distribution (normalized to 1) of coalescing binary BHs that can be detected with he current O3 LIGO/Virgo sensitivity calculated with an account of the star formation history in galaxies. The upper panel shows he results for the direct collapse  of the C-O stellar core into BH (without fallback from the envelope):  $M_{tot} = 0.9(M_{CO1}+M_{CO2})$. The bottom panel: model with the partial fallback of the envelope into BH, with the BH mass as in \cite{2012ApJ...749...91F}. 
The assumed common envelope efficiency  is 
$\alpha_\mathrm{CE} = 1.0$  and the core-envelope coupling time is  $\tau_c =5\times 10^5$ years.
The black symbols with error bars show the events from the published O1/О2 LIGO/Virgo Collaboration Catalogue.The color grade for events from  \cite{2019arXiv190210331Z},\cite{2019arXiv190407214V} (open triangles and circles) reflect the probability of their astrophysical origin.
}
\label{f:BHspins}
\end{figure}

The results of the population synthesis calculations of coalescing binary BHs with taking into account the rotational evolution of cores of massive stars are presented in Fig. 
\ref{f:BHspins} (see 
\cite{2019MNRAS.483.3288P} for more detail). 
This Figure shows the expected distribution of the effective spins 
$\chi_\mathrm{eff}$ of coalescing binary BHs that can be detected by LIGO/Virgo interferometers with the current O3 sensitivity
\footnote{In the ongoing O3 observations, the detection horizon of NS+NS binaries with a chirp mass of 1.2  $M_\odot$ averaged over the viewing angle is about 120 Mpc. The detection horizon for the coalescing compact binaries depends on the chirp mass as  $D_h\sim {\cal M}^{5/6}$ \cite{2010arXiv1003.2481T}.}, as a function of the total mass of the binary 
$M=M_{tot}$ 
(the upper left plot on each panel) for two BH formation models described above. The common envelope efficiency is assumed to be 
$\alpha_\mathrm{CE}=1$. The initial spin axes of the binary components are assumed to be randomly misaligned. The grey gradient color shows the probability of formation of the coalescing binary BHs convolved with the metallicity and star formation rate evolution with look-back time (redshift) in galaxies.    
Squares with error bars show the parameters of the observed binary BHs from the O1/O2 catalog \cite{LIGOO2}. 
The light-grey open circles (shown in green online) represent additional BH+BH binaries reported from independent O1/O2 LIGO/Virgo data analysis \cite{2019arXiv190407214V}. 
The lighter symbols correspond to higher probability of spurious (non-astrophysical) detections. 
It is seen that the results for the BH formation model without fallback (the upper panel) approximately describe the narrow distribution of the effective spins around zero (but for as yet unconfirmed sources found in  
\cite{2019arXiv190407214V}) (the upper right plot in the upper panel),
but do not encompass the most massive coalescing BHs (the bottom left plot on the upper panel). In the model of the BH formation with fallback from the rotating envelope (the bottom panel in Fig. \ref{f:BHspins}), the total masses of the coalescing binary BHs better correspond to observations and the effective spins of the coalescing binary BHs can be much larger, which have not yet been reliably found. Clearly, the successful performance of the LIGO/Virgo detectors during the O3 run started in April 2019 
\footnote{As of time of writing, about two dozen new BH+BH detections have been reported, see the list in https://gracedb.ligo.org/latest/. The parameters of these binaries will be reported after careful analysis only in the end of 2019.} will enable significant growth in the source statistics and more accurate comparison with models.

\begin{figure}
\includegraphics[width=\columnwidth]{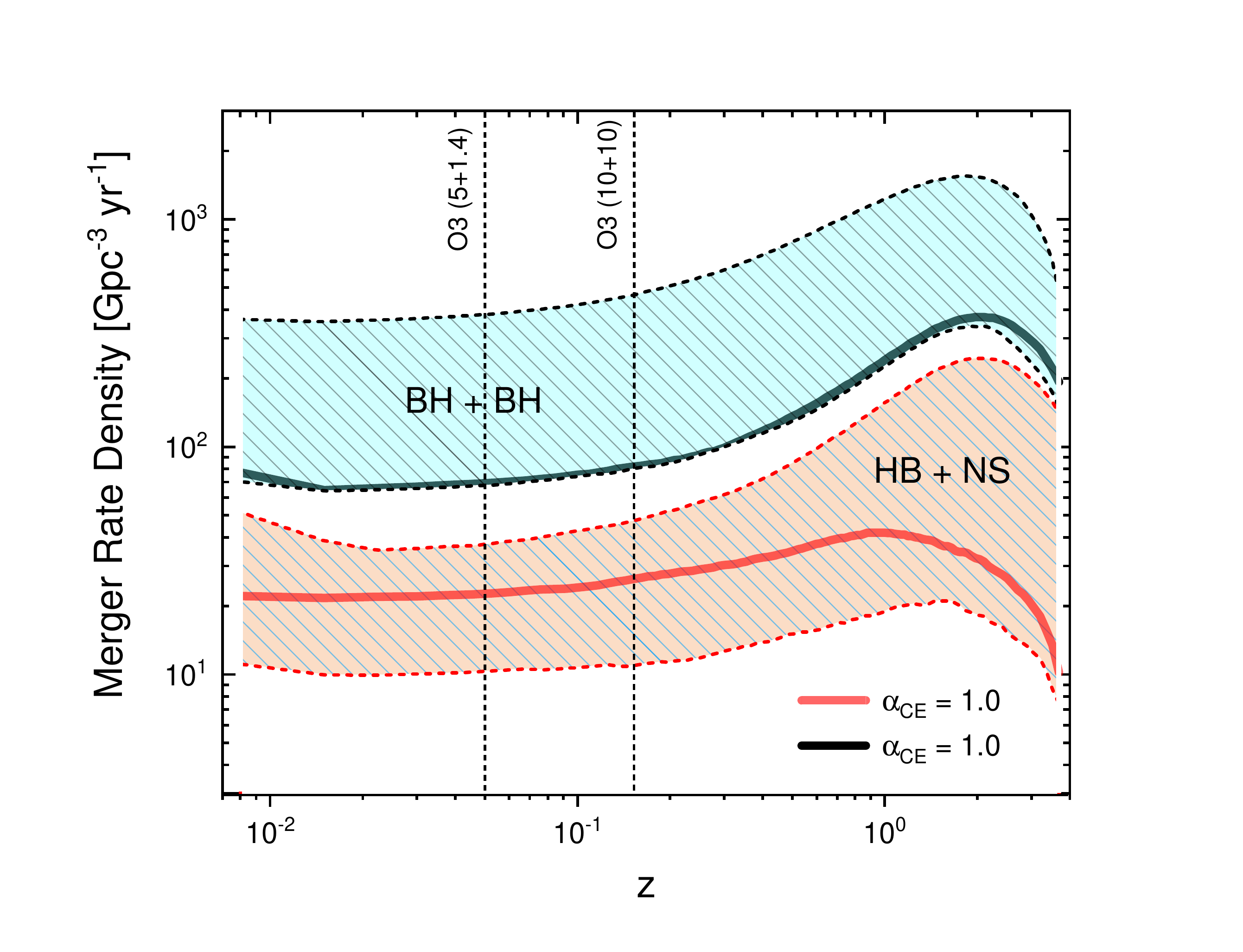}
\caption{Space density of coalescent rate of black hole -- black holes (BH+BH) and neutron star -- black hole (BH+NS) binaries (per year per cubic Gpc) as a function of cosmological redshift for a range of the common envelope efficiency parameter 
$\alpha_\mathrm{CE}$ 
with taking into account of the star formation rate and stellar metallicity evolution in the Universe. The upper and lower boundaries of the hatched regions correspond to 
 $\alpha_\mathrm{CE}= 4.0$ and $\alpha_\mathrm{CE}= 0.5$, 
 respectively, the thick lines correspond to $\alpha_\mathrm{CE}= 1.0$. The vertical dashed lines show the LOGO/Virgo O3 detection horizon for the coalescing compact binaries with component masses 
 $5+1.4 M_\odot$ and $10+10 M_\odot$. }
\label{f:BHrates}
\end{figure}

\begin{figure}
\includegraphics[width=0.9\textwidth]{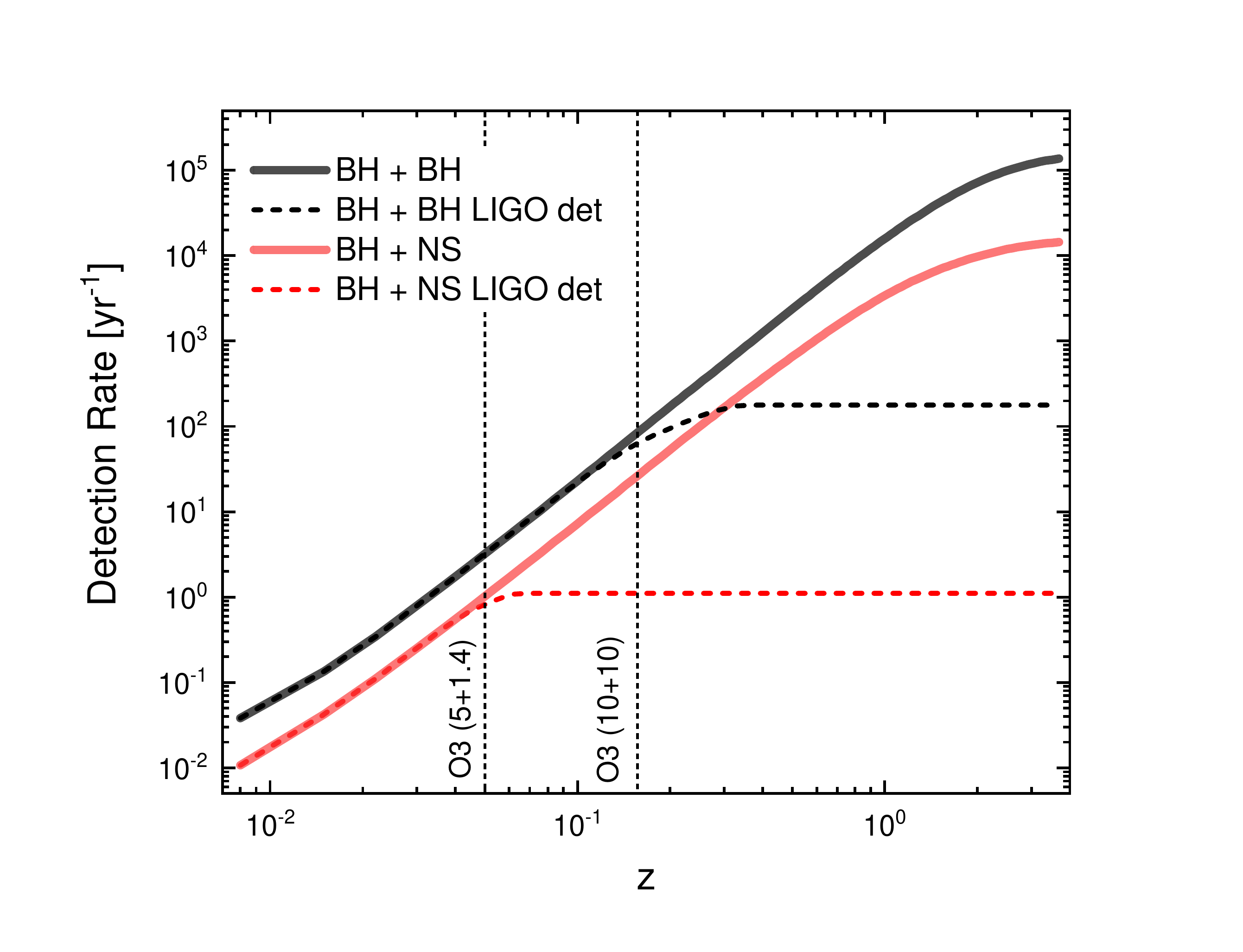}
\caption{The expected event rate per year of compact binary coalescences (integrated volume rate up to distances corresponding to given redshift z) as a function of the limiting redshift (the detection horizon). The rate is calculated with an account of the star formation and stellar metallicity evolution in the Universe and for the assumed common envelope efficiency parameter 
$\alpha_\mathrm{CE} = 1$. The thick black and red curves correspond to BH+BH and BH+NS coalescences, respectively. The dashed curve show the expected number of LIGO/Virgo O3 detections for the average orbit viewing angles ${\cal R}_\mathrm{BHBH}\sim 2\times 10^2~\hbox{yr}^{-1}$ and ${\cal R}_\mathrm{BHNS}\sim 1~\hbox{yr}^{-1}$
for BH+BH and BH+NS events, respectively, 
The vertical dashed lines show the O3 LIGO/Virgo detection horizon for binaries with masses $5+1.4 M_\odot$ and $10+10 M_\odot$.
 }
\label{f:BHrates2}
\end{figure}

\begin{figure}
\includegraphics[width=0.9\textwidth]{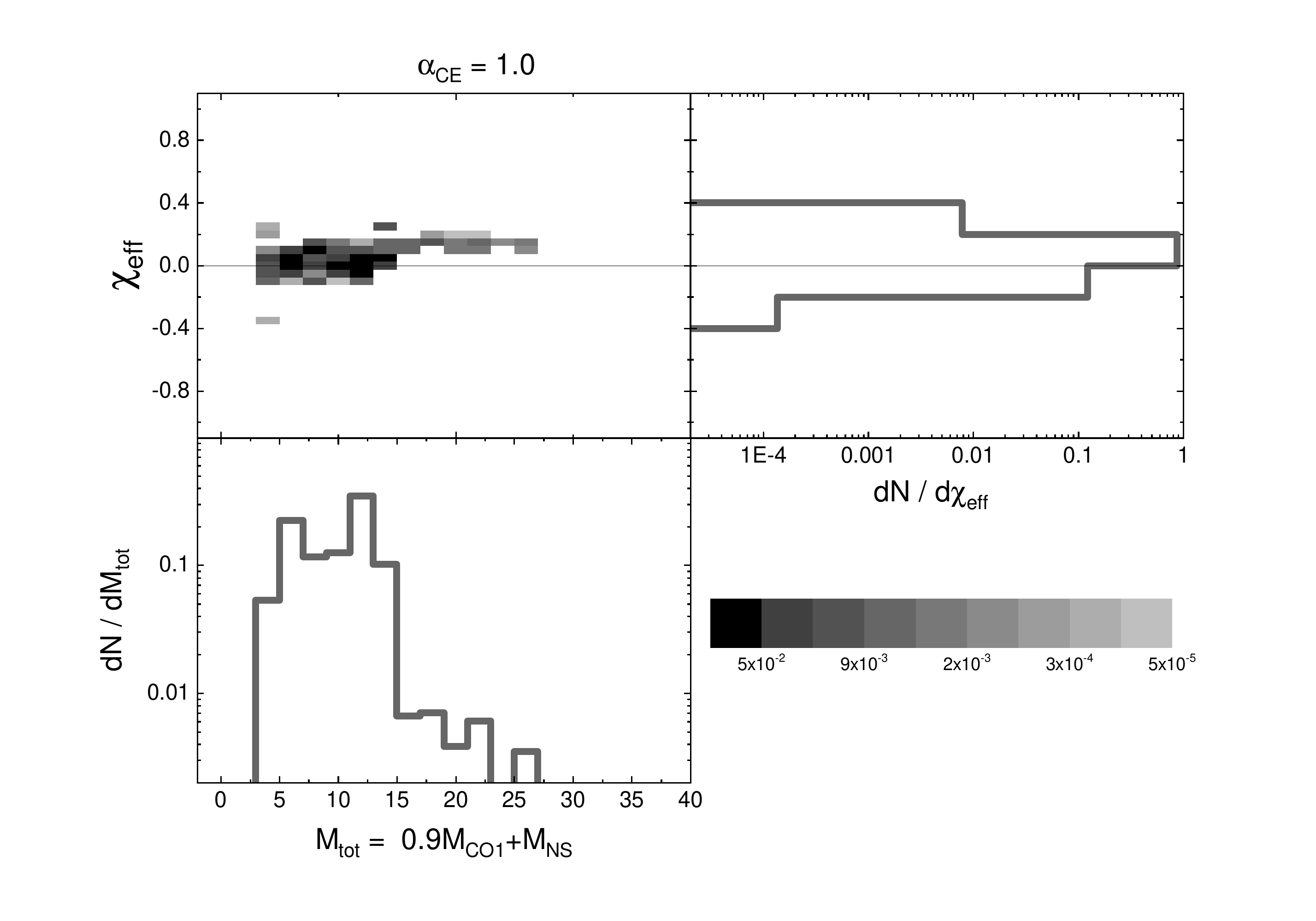}
\includegraphics[width=0.9\textwidth]{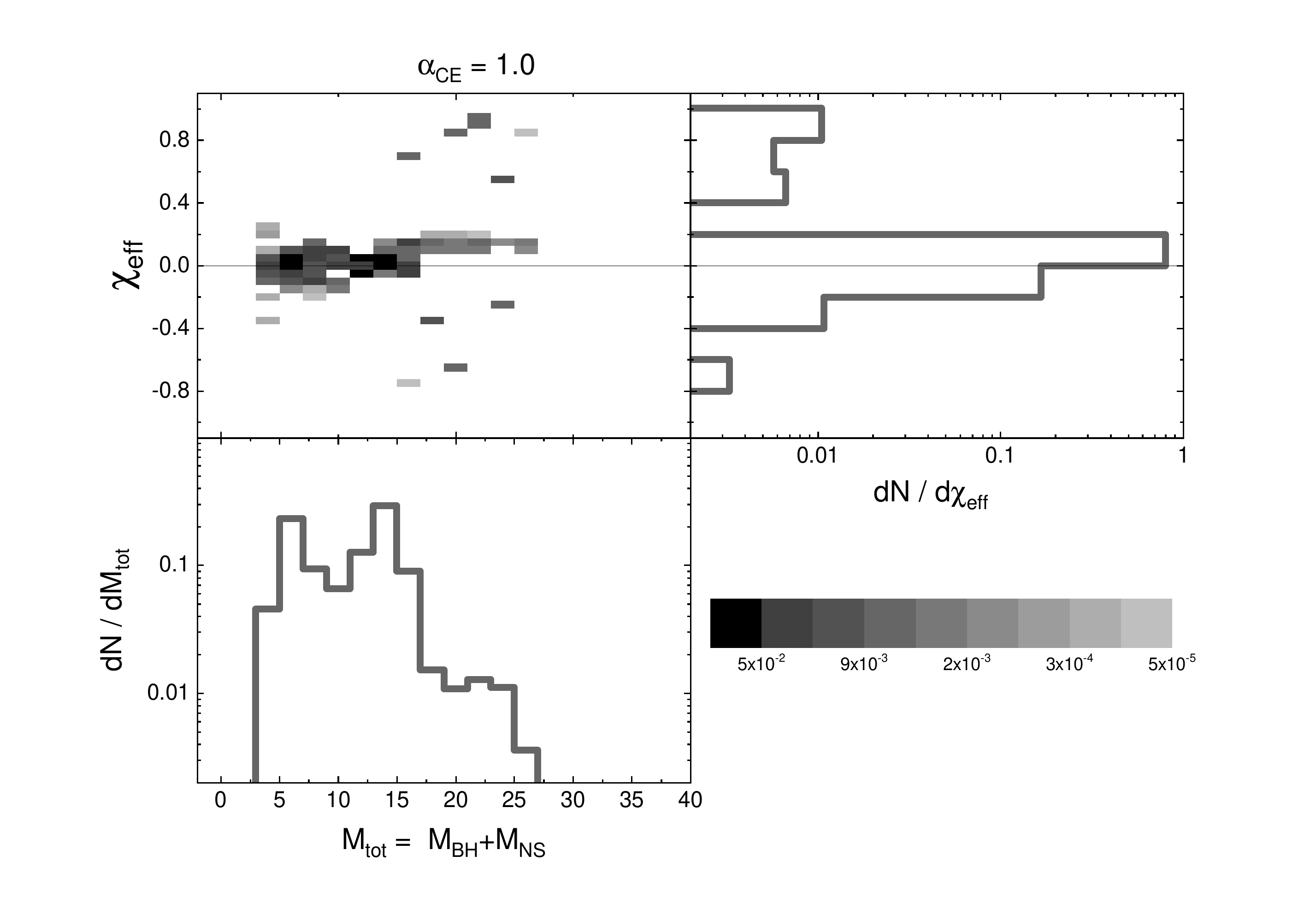}
\caption{The model effective spin and total mass distributions for the coalescing BH+NS binaries. The BH formation parameters are the same as for BH+BH binaries in Fig. \ref{f:BHspins}. }
\label{f:BH_NS_spins}
\end{figure}

\section{Spins of coalescing BH+NS binaries}
\label{s:nsbh}

The widely recognized evolutionary model of massive binary stars \cite{2014LRR....17....3P} also predicts the 
formation of close binary systems containing neutron stars with black holes (BH+NS). The NS in such a binary observed as a radio pulsar would offer the unique probe of space-time around the companion BH by accurate pulsar timing measurements. Population synthesis calculations have predicted about one PSR+BH system per several thousand single pulsars in the Galaxy
\cite{1994ApJ...423L.121L}, but searched for such binaries have failed so far. There is hope that such binaries can be first discovered by gravitational astronomy methods. The formation rate of BH+NS binaries is much lower than that of NS+NS and BH+BH binaries because the formation of the secondary (lighter) NS companion in a massive binary is accompanied with significant mass loss from the system and additional kick imparted to NS during the supernova explosion, which in many cases can result in disruption of the binary system (see e.g.  \cite{2001PhyU...44R...1G,2014LRR....17....3P} for more detail). 

Fig. \ref{f:BHrates} shows the calculated space density of binary BH+BH and BH+NS coalescences (per year per cubic Gpc) for the model assumptions as in paper 
\cite{2019MNRAS.483.3288P} on the evolution of the stellar metallicity and star formation rate in the Universe. 
The common envelope efficiency varied in the range 
$0.5\le \alpha_\mathrm{CE}\le 4$. The NS kick velocity is assumed to follow a Maxwellian distribution with the mean value 256 km/s. It is seen that the coalescence rate of BH+NS binaries per unit volume is at least by an order of magnitude smaller than that of binary BH+BH, which so far has not contradicted to the observed LIGO/Virgo detection statistics.  

Fig. \ref{f:BHrates2} presents the expected model detection rate in as a function of the limiting redshift (the integrated coalescence rate per unit volume up to the distance corresponding to a given $z$). The solid lines show the detection rate of binary black hole (BH+BH) coalescences with an account of the star formation rate history for the common envelope efficiency parameter $\alpha_\mathrm{CE} = 1$.
The dashed lines marked with 'BH+BH LIGO det' and 'NS+BH LIGO det' show the expected LIGO/Virgo O3 detection rate 
${\cal R}_\mathrm{BHBH}\sim 2\times 10^2~\hbox{yr}^{-1}$ and ${\cal R}_\mathrm{BHNS}\sim 1~\hbox{yr}^{-1}$ of BH+BH and BH+NS coalescences, respectively. The specific volume and detection rates of BH+NS coalescences shown in Fig. \ref{f:BHrates} and \ref{f:BHrates2} are in agreement with independent calculations
(see, for example,  \cite{2019arXiv190604197B}).

Fig.  \ref{f:BH_NS_spins} shows the effective spin distributions of coalescing BH+NS binaries  \eqn{e:chieff} as a function of the total mass for two BH formation models 1) and 2) (as in Fig.  \ref{f:BHspins}). 
The dimensionless NS spin is defined in the same way as for a BH:
$a^*_{NS}=J_{NS}/M_{NS}^2$, where $J=I\omega_{NS}=2\pi I/P_{NS}$ is the NS angular momentum ($I$ is the NS moment of inertia). The NS angular momentum was calculated with taking into account the evolution of NS rotational period $P_{NS}$ in a binary system (see \cite{2009ARep...53..915L} for more detail). 
Generally, the effective spin distribution of coalescing NS+BH 
binaries for different BH formation models is symmetric around tghe zero value but broader than that for binary BH+BH  (see Fig. \ref{f:BHspins}). 
Negative values of $\chi_\mathrm{eff}$ for BH+NS binaries are due to the randomly directed NS kick velocity. The total mass of the coalescing BH+NS binaries in these calculations does not exceed 
27 $M_\odot$. 
 
\section{Spins of coalescing primordial black holes}
\label{s:pbh}

\begin{figure}
	\includegraphics[width=0.9\columnwidth]{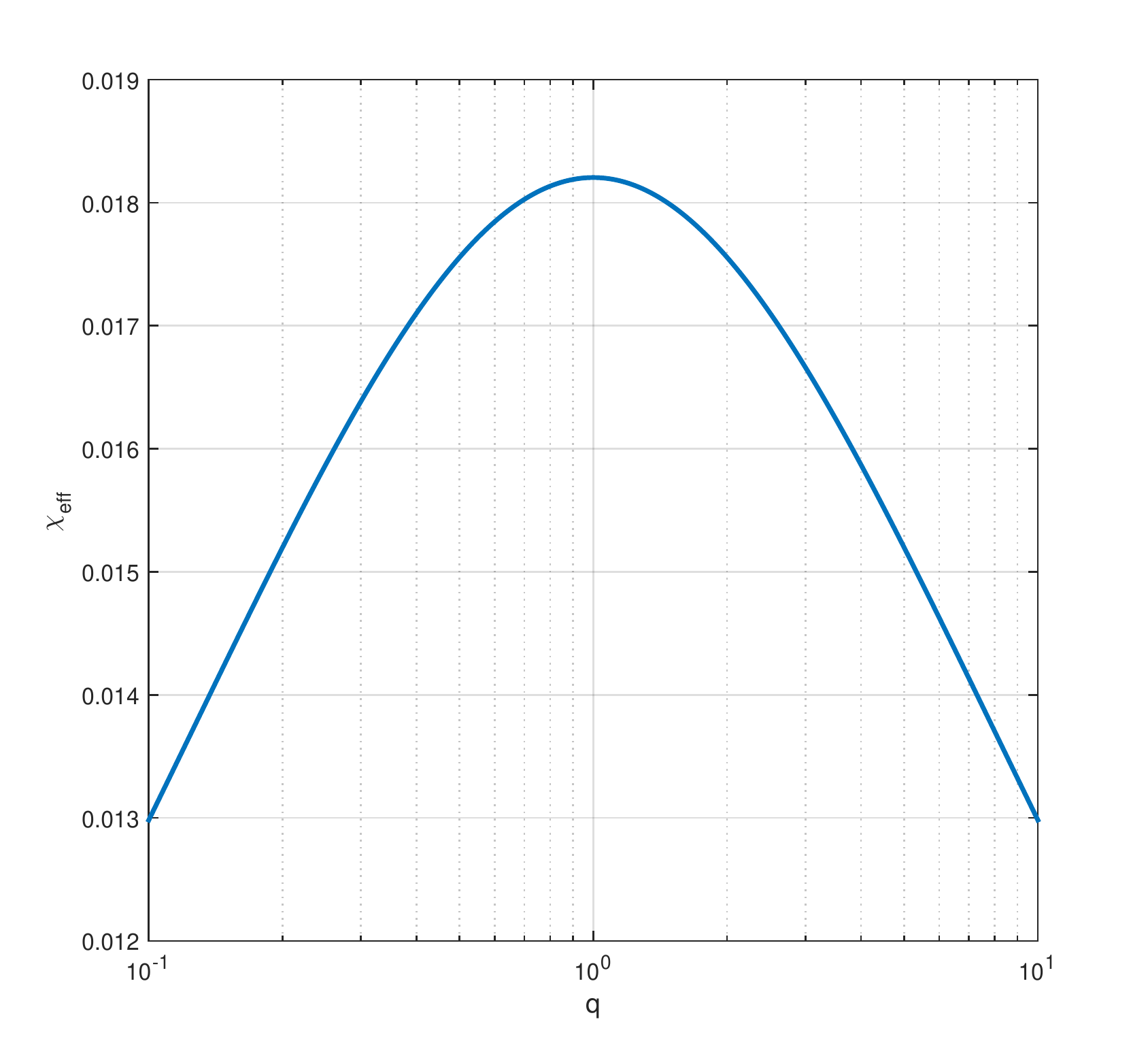}
    \caption{Maximum effective spin of a coalescing primordial binary BH with the chirp mass ${\cal M}=30 M_\odot$ acquired due to accretion in the medium with the sound velocity   $c_s=10^{-5}=3$~km~s$^{-1}$ and density $\rho=10^{-24}$~g~cm$^{-3}$ as a function of the binary component mass ratio $q$. Fig. from \cite{2019arXiv190400570P}.}
    \label{f:chimax}
\end{figure}

Near-zero effective spins of the coalescing binary BHs detected by LIGO/Virgo can be related to their origin and evolution and are widely discussed in the literature 
(see, for example,  
\cite{2017arXiv170607053B,2017PhRvD..96b3012T,2018PhRvD..98h3007N,2018arXiv180701336P,2019MNRAS.483.3288P,2019arXiv190513019F}
and references therein). As shown above, the narrow distribution of the effective spins near zero can be reproduced in the standard astrophysical channel of the binary BH formation from massive binary systems \cite{2016Natur.534..512B} assuming no additional fallback from the rotating stellar envelope 
(model 1) in Section  \ref{s:bhbh}, Fig. \ref{f:BHspins}, the upper panel). 
In addition to astrophysical scenarios of binary BH formation
\cite{2018arXiv180605820M} in which 10-50 $M_\odot$ BHs result from core collapse of massive stars, the possibility of the existence of the primordial binary BHs originated in the early Universe from primordial cosmological perturbations has been actively discussed. 
\cite{1997ApJ...487L.139N,2016PhRvL.116t1301B,2016PhRvD..94h3504C,2016JCAP...11..036B,2016PhRvL.117f1101S}. 
Spins of the primordial BHs should be nearly zero
(see \cite{2019arXiv190105963M,2019arXiv190301179D} for the recent discussion), and hence the effective spin of coalescing primordial binary BHs is expected to be small. 

To check the latter statement in paper
\cite{2019arXiv190400570P} we studied the possibility of accretion spin-up of BH components of a primordial binary BH in the external medium. 

An initially non-rotating BH acquiring mass  $\Delta M=M_f-M_0$ from the ambient medium through an accretion disc
gets the spin \cite{1970Natur.226...64B}:
\beq{e:spin}
a^*=\sqrt{\frac{2}{3}}\myfrac{M_0}{M_f}\left[4-\sqrt{18\myfrac{M_0}{M_f}^2-2}\right]\,
\eeq
(if no angular momentum is taken away by the disc radiation; the formula is valid for $M_f/M_0<\sqrt{6}$, otherwise $a^*=a^*_{max}\simeq 0.998$, the so-called Thorne limit
\cite{1974ApJ...191..507T}).
For $\Delta M\ll M_0$ from \eqn{e:spin} we obtain the estimate $a^*\simeq 9/\sqrt{6} (\Delta M/M_0)$.

For a single BH with mass $m$ moving with a velocity $v$ the fractional mass increase due to the Bondi-Hoyle-Lyttleton accretion over time  $t_0$ is
\beq{e:dMM}
\Delta M/M_0 \approx 4 \pi \rho m/(v^2+c_s^2)^{3/2}\times t_0, 
\eeq
where $\rho$ is the medium density, $c_s\sim \sqrt{T}\approx 10^{-5}\sqrt{T/1 \hbox{eV}}$ is the sound velocity in the medium with temperature $T$. For a 1 $M_\odot$ BH in the medium with  
density v$\rho\sim 10^{-24}$ g~cm$^{-3}$ under the condition $v\gg c_s$ over the Hubble time $t_0=t_H=4\times 10^{17} \mathrm{s}$ 
we obtain $\Delta M/M_0\simeq 1.7 10^{-3}m\ll 1$. 
Thus, the possible accretion-induced spin acquired by the single BH will be $a^*\simeq 3.76 \Delta M/M_0 \simeq 0.006 (m/M_\odot)$.
For BH with mass $m=30-50 M_\odot$ the spin can be noticeable but it is very difficult to measure the spin of a single BH. 

In a binary system, the effective BH spin can be measured from gravitational-wave observations. Integration of the accreting mass from the ambient medium over the orbit in the binary BH system 
\cite{2019arXiv190400570P} shows that for the two point-like masses $m_1$ and $m_2$ ($q = m_1/m_2$ is the binary mass ratio, $M = m_1 + m_2=m_1(1+1/q)$ is the total mass) the accretion-induced fractional mass growth of the mass $M_1$ over the Hubble time $t_H$ is 
\beq{e:dM1M1}
\left.\frac{\Delta M_1}{M_1}\right|_0=\dfrac{5\pi\rho M^{1/2}a_0^{11/2}}{88m_2^4}=\dfrac{5}{88}\bigg(\dfrac{256}{5}\bigg)^{11/8}\pi\rho t_H^{11/8}m_1^{5/8}q^{3/4}(1+q)^{15/8}.
\end{equation}
Here the well-known relation between the coalescence time of a binary due to GW emission with the initial orbital semi-major axis $a_0$ is used: $t_0=\dfrac{5a_0^4}{256Mm_1m_2}$. Numerically, for $m_1=30 M_\odot$ we get a very low value $(\Delta M_1/M_1)\sim 10^{-6}-10^{-7}$. 

However, the estimate of the accretion-induced BH mass growth in a binary system significantly increases is the initial orbit was highly eccentric, 
  $e_0\sim 1$. In this case \cite{2019arXiv190400570P}
\begin{eqnarray}
\label{e:dMM1}
\left.\frac{\Delta M_1}{M_1}\right|_e &\approx 10^{-5}
\myfrac{\rho}{10^{-24} \mathrm{g\,cm}^{-3}}\myfrac{M_1}{30 M_\odot}^{5/8}
q^{3/4}(1+q)^{15/8}\myfrac{0.1}{1-e_0^2}^{2.58}\nonumber \\
&\approx 10^{-5}
\myfrac{\rho}{10^{-24} \hbox{g\,cm}^{-3}}\myfrac{\cal M}{30 M_\odot}^{5/8}
q(1+q)^2\myfrac{0.1}{1-e_0^2}^{2.58}\,.
\end{eqnarray}
In the last equation the chirp mass of the binary system was introduced, ${\cal M}\equiv (M_1M_2)^{3/5}/M^{1/5}=M_1(q^2(1+q))^{-1/5}$, which can be directly inferred from the analysis of the spiraling GW waveforms from a coalescing binary consisting of two point-like masses.

The effective spin of a primordial binary BH accreting from the external medium is found to be 
\begin{eqnarray}
\label{e:chieffpbh}
\chi_{\mathrm{eff}}&=\frac{q}{1+q}a^*_1+\frac{1}{1+q}a^*_2\approx 
3.76\times 10^{-5}\myfrac{\rho}{10^{-24} \mathrm{g\,cm}^{-3}}\myfrac{\cal M}{30 M_\odot}^{5/8}
\myfrac{0.1}{1-e_0^2}^{2.58}(1+q)(q^2+q^{-3})\nonumber\\
&>5.3\times 10^{-4}\myfrac{\rho}{10^{-24} \mathrm{g\,cm}^{-3}}\myfrac{\cal M}{30 M_\odot}^{5/8}
\myfrac{0.1}{1-e_0^2}^{2.58}
\end{eqnarray}
for any mass ratio $q$
because the function $f(q)=(1+q)(q^2+q^{-3})$ has minimum  $f(q_{min})=4$ at $q_{min}=1$. 
An account for restriction for maximum possible value of the initial eccentricity for the given sound velocity of the external medium  $c_s$ leads to the maximum possible 
$\chi_{\mathrm{eff}}$ shown in Fig. \ref{f:chimax}. 
A rough upper limit can be written as 
$$
\chi_{\mathrm{eff},\mathrm{max}}\simeq 0.01(\rho/10^{-24}\hbox{g\,cm}^{-3})({\cal M}/30\,M_\odot)^{0.97}(c_s/10^{-5})^{-2.75}
$$ 
for a wide mass ratio interval $0.1<q<10$. This estimate suggests that the effective spin of coalescing primordial binary black holes can be a few percents due to accretion of gas from cold interstellar medium. In this spin-up mechanism, the spins of both BH components must be aligned with the binary orbital angular momentum.  

\section{Conclusion}

Measurements of BH rotation from astronomical observations remain actual problem of modern astrophysics. The angular momentum (spin) of a BH is the second important (after the mass) parameter determining the structure of space-time around the BH. The rotation of BHs in close X-ray binaries and in galactic nuclei is likely to be the necessary condition for launching relativistic jets in AGNs and microquasars and can determine the structure of the outflows (see the review \cite{Beskin:2010}). 
Gravitational-wave astronomy offered new possibilities of direct measurement of BH spins in coalescing binary BHs. The current LIGO/Virgo results 
\cite{LIGOO2} suggest a fairly narrow effective BH spin distribution around zero $\chi_\mathrm{eff}\sim 0$, which can be used to understand the origin of coalescing binary BH with masses 10-60 $M_\odot$ \cite{2018ApJ...868..140T,2018PhRvD..98h4036G}.

We have shown (see \cite{2019MNRAS.483.3288P} and Fig.  \ref{f:BHrates})
that the astrophysical channel of coalescing binary BH formation from the evolution of massive binary stars
with taking into account of the evolution of stellar metallicity and star formation rate in the Universe leads to the effective BH spin and total mass of coalescing binary BHs distribution that does not contradict to the current LIGO/Virgo results \cite{LIGOO2}. 
Additional fallback of matter from the rotating stellar envelope onto a newborn BH resulted from the massive star iron core collapse can significantly spin up the BH leading to a wide effective spin range (see lower panel in Fig. 
\ref{f:BHspins}). 
The rapid rotation of components of coalescing binary BHs has not yet been reliably found
\cite{LIGOO2}, although there are indications that such
binaries are present in the O1/O2 LIGO/Virgo data
\cite{2019arXiv190210331Z,2019arXiv190407214V}. These additional sources require confirmation.

Undiscovered up to now (June 2019) remain coalescing compact binaries hosting a black hole and a neutron star. 
The existence of such systems firmly follows from the modern theory of evolution of massive binaries. 
Using the same model assumptions of the BH formation, we have calculated the coalescence rates of BH+NS binaries (Fig. \ref{f:BHrates}) and their detection rate by the O3 LIGO/Virgo interferometers (Fig. \ref{f:BHrates2}).
For the standard assumptions on the binary evolution parameters the detection rate of BH+NS binaries is by more than two orders of magnitude lower that that of BH+BH binaries (see the dashed curves in Fig. 
 \ref{f:BHrates2}): 
 ${\cal R}_\mathrm{BHBH}\sim 2\times 10^2~\hbox{yr}^{-1}$  и ${\cal R}_\mathrm{BHNS}\sim 1~\hbox{yr}^{-1}$. 
 Chances of detection of these type of coalescing compact binaries in the ongoing O3 LIGO/Virgo observations are not very high. We also calculated the expected distribution of the effective spins of BH+NS binaries depending on the assumed BH formation model.
(Fig. \ref{f:BH_NS_spins}). 
Due to the substantial kick velocity imparted to a newborn NS the effective spins of BH+NS coalescing binaries fall within a wide range $-0.4<\chi_\mathrm{eff}<0.4$ even in the conservative case without fallback of matter from the rotating stellar envelope onto newborn BH (the upper panel in Fig. \ref{f:BH_NS_spins}).

We have also considered the possible spin-up of primordial BHs due to gas accretion from cold interstellar medium. We have shown (see \cite{2019arXiv190400570P} and Fig. \ref{f:chimax}) that the accretion spin-up of primordial BH in binary systems can result in the effective spins at several percent level, therefore small positive effective spins of the observed coalescing BH+BH does not contradict the hypothesis of their possible primordial origin.  

\section*{Acknowledgements}

The work of KAP is partially supported by RSF grant
19-42-02004 (Sections 1.1-1.3, 4.2.2 and the general supervision). AGK and NAM acknowledge the support by the Scientific School of M.V. Lomonosov Moscow State University 'Physics of Stars, relativistic Objects and Galaxies'.

\bibliographystyle{UFN}
\bibliography{bhs}
\end{document}